# Modeling liquefaction-induced runout of a tailings dam using a hybrid finite element and material point method approach

Brent Sordo[i], Ellen Rathje, Ph.D.[ii], Krishna Kumar, Ph.D.[iii]

i) Ph.D. Candidate, Department of Civil, Architectural, and Environmental Engineering, The University of Texas at Austin, 301 E Dean Keeton St, Austin, TX 78712; E-mail: bsordo@utexas.edu
ii) Professor, Department of Civil, Architectural, and Environmental Engineering, The University of Texas at Austin, 301 E Dean Keeton St, Austin, TX 78712; E-mail: e.rathje@mail.utexas.edu
iii) Assistant Professor, Department of Civil, Architectural, and Environmental Engineering, The University of Texas at Austin, 301 E Dean Keeton St, Austin, TX 78712; E-mail: krishnak@utexas.edu

**ABSTRACT**

Tailings dams impound large amounts of saturated soil which can be highly susceptible to liquefaction. Liquefaction results in a severe loss of strength in the retained soil and potentially failure of the dam. If the dam is breached, a massive debris flow of liquefied soil is then released with potentially disastrous consequences downstream. Numerical models are frequently utilized to predict the liquefaction response of tailings dams and the potential runout, and these analyses inform engineering decisions regarding hazard avoidance and mitigation. The Finite Element Method (FEM) is a widespread tool which excels at modeling liquefaction triggering and initial movements, but it quickly loses accuracy when modeling large deformations due to mesh distortion. Conversely, the Material Point Method (MPM), a hybrid Eulerian-Lagrangian method, employs particles that move freely across a background grid and can account for large deformations without losing accuracy. However, issues with the accuracy of MPM's stress distributions and limits associated with the available boundary conditions impair its ability to predict liquefaction initiation. In this paper, we utilize a sequential hybridization of the FEM and MPM methods as a superior alternative to either individually. To demonstrate the efficacy of this hybrid method to simulate the entire process of tailings dam failures from initiation to runout, we model the 1978 Mochikoshi Tailings Dam failure. In this case, the dam collapsed during the main shaking of an earthquake due to the combined effects of the inertial seismic loading and liquefaction. We initiate this model in FEM to capture the immediate effects of the earthquake: the seismic response and liquefaction triggering. Then, we transfer the model into MPM, inheriting the FEM failure mechanism and capturing the runout behavior without mesh issues. The analysis successfully captures the liquefaction triggering and movements, but underestimates the final runout. Further refinements to the MPM phase of the analysis are required to better capture the large runout response.

## 1 INTRODUCTION

Mine tailings deposits are reservoirs of finely crushed waste rock from mining operations, typically impounded by massive earth embankments called tailings dams, and they represent some of the largest engineered structures in the world (Chambers and Higman, 2014). Due to the size of these structures, their failures can release large volumes of debris with extensive runouts and severe consequences to downstream populations (Yang et al., 2022) and, as they can often contain harmful chemicals, the environment (Kossoff et al., 2014). The severity of these disasters is difficult to understate, yet the reliability of tailings dams is among the lowest of earth structures (Psarropoulos and Tsompanakis, 2008). In 2015, the unexpected Fundão Tailings Dam Failure of the Samarco mine in Brazil became the worst environmental disaster in that country's history (Laurrari and Lall, 2018). Engineers are thus tasked with the critical task of protecting downstream communities and environments by accurately evaluating the stability of these dams and estimating the extent of the runout in the event of failure.

Tailings are transported as a slurry (Laurrari and Lall, 2018), and thus the sand to silt sized particles form loose and saturated deposits that are highly susceptible to liquefaction during earthquakes (Ishihara, 1984). As such, earthquakes are among the three most frequent causes of tailings dam failures (ICOLD, 2001). Liquefaction greatly reduces the shear resistance within the tailings and, potentially combined with the inertial effects of the inertial seismic loading, can result in catastrophic dam failures (Ishihara, 1984). This was the cause of the failure of two Japanese tailings dams in the 1978 Izu-Ohshima-Kinkai earthquake, two Chilean tailings dams in the 1985 Algarrobo earthquake, a Peruvian tailings dam in the 1996 Nazca earthquake, and many other incidents (Psarropoulos and Tsompanakis, 2008).

The rapid development of numerical modeling has revolutionized the ways engineers estimate the seismic response of dams, the effect that response has on stability, and the runout potential (Chakraborty and Choudhury, 2009). The Finite Element Method (FEM) is among the most common numerical methods for modeling the seismic and pore pressure responses of tailings dams (Chakraborty and Choudhury, 2009) and can capture the initiation of failure well, but it quickly loses accuracy when modeling large-deformations due to mesh distortion and eventually, mesh entanglement. As a result, it is critically restricted in its ability to model the runout of tailings dam failures (Soga et al., 2016; Sordo et al., 2023) which can involve displacements on the order of kilometers.

The Material Point Method (MPM; Sulsky et al., 1994; Bardenhagen et al., 2000), a hybrid Eulerian-Lagrangian method, is an alternative numerical approach suitable for modeling large deformation problems such as the runout of tailings dam failures. However, while MPM is numerically capable of accounting for large displacements without losing accuracy (Soga et al., 2016), it has shortcomings associated with computing accurate stresses at moving material points due to cell crossings (Bardenhagen and Kober 2004) and with applying the absorbing boundary conditions required for seismic analysis (Alsardi et al. 2021). These issues limit the accuracy of MPM predictions of failure initiation and thus, ultimately, its final runout predictions (Sordo et al., 2023). Consequently, while numerical methods currently exist that can effectively determine if a tailings dam failure will occur, current methods lack the ability to accurately predict that failure's associated runout and engineers still often rely on empirical predictions for runout volume and extent (Laurrari and Lall, 2018).

To overcome the shortcomings of each method, we employ a novel hybrid FEM-MPM method (Sordo et al., 2023) in which a single model simulates the entire failure process from initiation (via FEM) through runout (via MPM). The hybrid approach combines the advantages of both numerical methods to comprehensively evaluate the slope failure process. The hybrid FEM-MPM models begin with an "initiation phase" or "FEM phase", in which the initiation mechanism and the initial inelastic movements are captured by FEM. Then, at a user-specified time, the "transfer phase" is performed in which the state of the FEM model is converted into an MPM model. This begins the "runout phase" or "MPM phase" in which the model is allowed to runout into its final configuration. While FEM and MPM have been hybridized before, only recently have they been sequentially hybridized over the same model (Sordo et al., 2023). This hybrid method has been applied to simulations of simple, gravity-driven slope failures (Sordo et al., 2023), but it has not yet been utilized to its full potential to comprehensively model seismic liquefaction induced slope or dam failures.

The Mochikoshi Tailings Dams were three tailings dams which retained the Hozukizawa disposal pond (Figure 1), two of which failed during the 1978 Izu-Ohshima-Kinkai earthquake in Japan (Ishihara, 1984). These are prime examples of seismic-liquefaction induced tailings dam failures. We select the failure of the No. 1 dam as a case study to model and demonstrate the full capability of the hybrid FEM-MPM procedure.

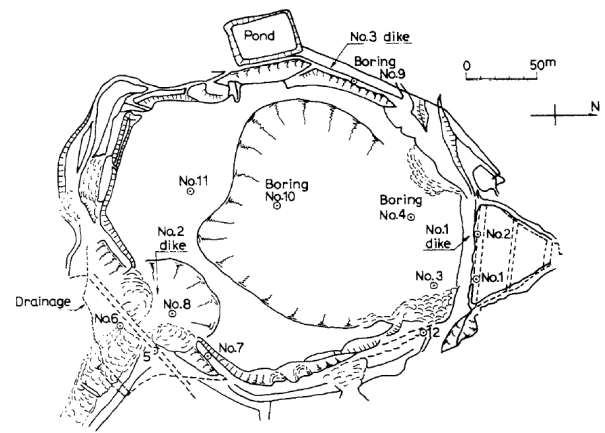

Fig. 1: Plan view of Hozukizawa disposal pond after failures (Ishihara, 1984).

On the morning of January 14, 1978, the Izu-Ohshima-Kinkai earthquake (M=7.0) shook the site (Okusa and Anma, 1980). The No. 1 dam failed during the main shaking, and the No. 2 dam failed the next day after an M=5.8 aftershock (Ishihara, 1984). The No. 3 dam did not collapse. Sand boils found at the site in subsequent investigations confirmed that the retained tailings had liquefied (Okusa and Anma, 1980). A witness to the failure of the No. 1 dam testified that the upper dam swelled and then breached within 10 seconds after the main shock (Ishihara, 1984). An investigation of the causes of these failures by Ishihara (1984) concluded that the No. 1 dam breach was caused by the combined effects of inertial seismic loading and liquefaction in the tailings, whereas the No. 2 dam failed five hours after an aftershock due to pore pressure redistribution via fissures in the embankment created by the shaking.

Ishihara (1984) created a limit equilibrium model of the No. 1 dam failure and confirmed that the failure was caused by a combination of the effects of liquefaction and inertial seismic loading. Later, Byrne and Seid-Karbasi (2003) performed nonlinear finite element analyses of the No. 1 dam and successfully captured the triggering and initial deformation patterns of the failure. We elect to model this dam failure with our hybrid FEM-MPM method as it features the combined effects of liquefaction and inertial seismic loading as well as an extensive (~800-meter; Byrne and Seid-Karbasi, 2003) runout. Thus, the triggering mechanism will rely on the constitutive precision and boundary conditions of FEM, and the runout will rely on the large-displacement capabilities of MPM. The need for the capabilities of both FEM and MPM make this failure an ideal demonstration of the hybrid FEM-MPM method.

## 2 GEOTECHNICAL AND GEOLOGICAL CONDITIONS OF THE MOCHIKOSHI DAM

The Hozukizawa disposal pond was constructed at the top of a mountain at the Mochikoshi site. The mountain was composed of volcanic tuff, the weathered surface of which was removed prior to the construction of the starter dam. The permeability of the mountain deposit was deemed sufficient to drain the tailings themselves, so the only drainage system created was at the toe of the starter dam (Ishihara, 1984).

The tailings deposit, at the time of the earthquake, was approximately 30 meters deep and consisted of thin, alternating layers of silt and sandy silt ranging in thickness from 3 to 7 cm. These tailings were deposited at a rate of approximately 2 m per year, pumped as a slurry from the mill along the river at the base of the mountain (Ishihara, 1984). Ishihara (1984) estimated the average $(N_1)_{60-cs}$ value of the tailings to be 6 and their permeability to be $10^{-4}$ cm/s and $10^{-7}$ cm/s in the horizontal and vertical directions, respectively. The anisotropy of the permeability was due to the stratified nature of the tailings deposit. Static triaxial tests conducted on undisturbed samples yielded Mohr-Coulomb effective strength parameters of c' = 0 and ϕ' = 30-39° (Ishihara 1984).

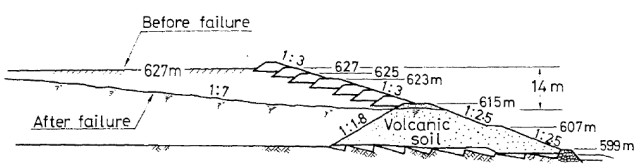

Fig. 2: The geometry of Mochikoshi Dam No. 1 (Ishihara, 1984).

To retain the tailings, dams were constructed along the perimeter of the pond using the upstream method. The base dams were constructed with compacted local soils ranging from silts to gravels of volcanic origin (Ishihara, 1984). The upper dams were created according to the upstream filling method as tailings were added over the next decade. Figure 2 shows the geometry of Mochikoshi Dam No. 1 at its maximum height. Based on static triaxial tests of undisturbed samples, the Mohr-Coulomb effective strength parameters of the embankment dam were estimated to be c' = 25 kPa and ϕ' = 35° (Ishihara 1984). The embankment dam material did not liquefy during the earthquake (Ishihara 1984).

## 3 HYRBID FEM-MPM PROCEDURE

The sequential hybrid FEM-MPM procedure (Sordo et al., 2023) consists of a three-phase workflow: (a) an initial FEM phase to capture initiation, (b) a transfer phase which transfer the soil state from FEM to MPM, and (c) a final runout phase in MPM.

The hybrid FEM-MPM models initialize in FEM with linear elastic constitutive models for every material present, creating a geostatic initial stress state. The constitutive models are then changed to elasto-plastic models that are capable of capturing failure initiation, and the earthquake loading is applied. This initial FEM phase progresses until a user-specific transfer time ($t_T$), at which the transfer phase is conducted.

The ideal choice for $t_T$ is not always obvious, but for a liquefaction problem the transfer must be conducted after the earthquake-induced pore pressure response is captured, as the MPM phase is incapable of accurately capturing this phenomenon. However, later transfer times will result in a loss of accuracy, because the FEM phase will have deteriorated due to mesh distortion, and any inaccurate developments in this phase will persist into the MPM phase and the final results. Thus, the transfer must also be performed before excessive mesh distortion or other unrealistic developments occur in the FEM phase. Judgement must be exercised by the modeler to idealize the balance between these two constraints on $t_T$.

The transfer phase consists of an algorithm which converts each FEM element into multiple material points which receive state variables based on their associated FEM nodes and elements. The material points are created at the Gauss locations of the element, although the number of material points does not need to be equal to the number of Gauss points in the FEM analysis, so their locations can differ. Some of the state variables in the FEM are stored at the Gauss points (stresses, strains, etc.) and some are stored at the nodes (position, velocity, etc.). The interpolation procedures used to define the state variables at the material point locations are described in Sordo et al. (2023). Each material point is also assigned a portion of the mass and volume of its respective element proportional to the Gauss weight of their respective locations.

Finally, the MPM phase is initiated from the particles whose initial conditions have been determined by the transfer algorithm. The MPM phase disregards the FEM mesh and instead utilizes a structured grid of square cells. In this final phase, the MPM analysis progresses until the runout process is complete, delivering the final runout results.

## 4 FEM PHASE OF MOCHIKOSHI MODEL

We perform the FEM phase of our hybrid FEM-MPM analysis of the Mochikoshi Tailings Dam No. 1 in the open-source FEM code OpenSees (McKenna, 1997). We obtain the dimensions of the dam from Ishihara (1984; Figure 2) and construct a partially structured mesh of quadrilateral elements in GiD (Ribó et al., 1999; Figure 3). The coarseness of the mesh varies, being the finest at and immediately surrounding the dam itself. The tailings are composed of a structured mesh, but the embankment itself requires an unstructured mesh due to its irregular geometry. We employ the SSPquadUP element (McGann et al., 2012) across the model, which is a reduced-order, linear, quadrilateral element with four nodes and a single central Gauss point. In our previous work (Sordo et al., 2023), we experimented with multiple element types in the FEM phase of the hybrid FEM-MPM procedure and found that this element is the most computationally efficient while maintaining accuracy.

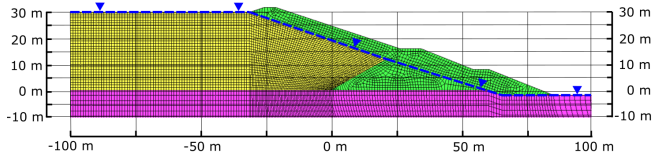

Fig. 3: Mesh of the FEM phase of the model of Mochikoshi Dam No. 1. Green indicates the embankment, yellow the tailings, and purple the foundation. Blue indicates the phreatic surface under geo-static conditions. Note that the mesh extends further left, right, and downward.

The FEM model is divided into three materials: the foundation, the tailings, and the embankment. For all three, we utilize the PM4Sand constitutive model (Boulanger and Ziotopoulou, 2015). The foundation and embankment soils were assigned the default PM4Sand properties (Boulanger and Ziotopoulou, 2015) for a dense and medium sand, respectively.

The tailings are treated as a loose sandy material, but the properties are modified from the default of Boulanger and Ziotopoulou (2015), as the default parameters greatly overestimated the undrained residual shear strength ($s_R$). Accurately modeling the residual shear strength is critical to accurately modeling the runout of the tailings, so the default properties were calibrated using a method described by Boulanger and Ziotopoulou (2018). This method is to shift the critical state line (CSL) to obtain a desired relation between undrained residual shear strength and effective stress and modify the contraction rate to maintain the target CRR curve of the default loose PM4Sand model. The $s_R$ values are calibrated to the median $s_R$ predicted by the empirical relation of Weber (2015). The $s_R$ values were derived from the tailings' $(N_1)_{60\text{-cs}}$ of 6, and ranged from $s_R$ = 3.2 kPa at an effective stress of 17 kPa to $s_R$ = 9.1 kPa at an effective stress of 186 kPa. The properties for each of the three materials are shown in Table 1.

Table 1: PM4Sand Parameters of Materials in Mochikoshi Model

|  |  | Tailings | Embankment | Foundation |
|---|---|---|---|---|
| Mass Density (ρ) (kg/m³) | | 1870 | 1690 | 1900 |
| Relative Density (Dr) | | 35% | 55% | 75% |
| Shear Modulus Coefficient ($G_0$) | | 476 | 677 | 890 |
| Contraction Parameter (hpo) | | 2.7 | 0.4 | 0.63 |
| CSL Parameters | Q | 11.998 | 10 | 10 |
| | R | 3.75 | 1.5 | 1.5 |

No time histories were recorded at or near the Mochikoshi site, so a recording from the 1971 San Fernando earthquake from Caltech's Athenaeum Library (N00E component from NGA-West RSN 79) is used, which is similar to the time history used by Byrne and Seid-Karbasi (2003) to analyze their FEM model of the dam. The time history is scaled to a PGA of 0.2 g, which is approximately the PGA what was estimated by Ishihara (1984) for the site. The earthquake time history is applied to the site as a velocity restraint upon the nodes at the base of the model which imposes horizontal movement and prevents vertical movement. The model boundary also features a Lysmer-Kuhlemeyer (1969) dashpot with properties to simulate bedrock.

Figure 4 displays vectors of the nodal velocities over the deformed FEM mesh at $t$ = 16 s, which represents the end of the strongest shaking. The FEM phase of the Mochikoshi failure delivers a failure mechanism similar to the forensic evidence of the failure initiation. The crest of the upper dam begins to sink while its toe begins to bulge, matching the witness statement that the upper dam swelled at the onset of the failure.

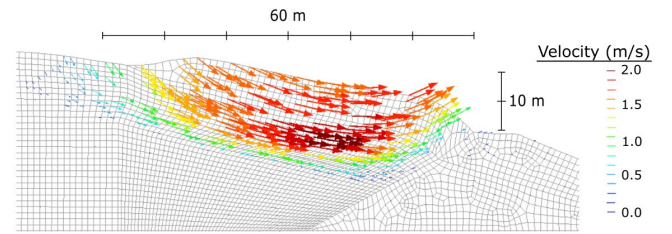

Fig. 4: FEM velocity vectors at $t$ = 16 s.

The failure surface predominantly cuts through the tailings underlying the upper dam, extending from crest to toe. The lower dam appears to remain structurally intact, forcing the tailings to pass over it, as was the case in the actual Mochikoshi failure. This is similar to th FEM response was by Byrne and Seid-Karbasi (2003).

The initial shaking ($t < 5$ s) generates a minimal amount of pore pressure in the tailings, but pore pressures in the tailings grow most rapidly between $t = 5$ s and 15 s, during the most powerful shaking. Figure 5 displays the contours of excess pore pressure ratio ($r_u$) at $t = 16$ s.

$$r_u = \Delta u / \sigma'_{vo} \quad (1)$$

At this time, the tailings up to a depth of approximately 12 m have reached an $r_u$ of at least 0.7, with the highest $r_u$ values located below the dam.

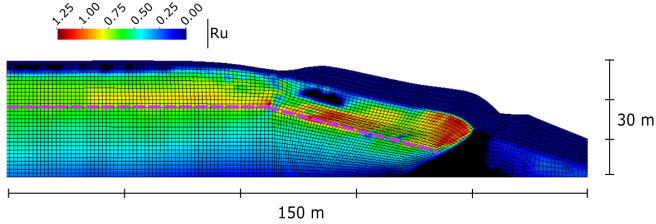

Fig. 5: $r_u$ contours at $t = 16$ s from FEM phase with zone of liquefaction approximately outlined in purple.

While the FEM phase successfully models the development of a failure surface and pore pressure response, it suffers from the typical limitations of FEM as the mesh deforms and entangles at $t \sim 22.5$ s. FEM results beyond the point of mesh entanglement are unreliable (Sordo et al., 2023), and therefore, FEM is unable to provide meaningful runout predictions.

## 5 TRANSFER PHASE OF THE MOCHIKOSHI MODEL

To optimally model the tailings runout, we must select an ideal time to facilitate the transfer from FEM to MPM ($t_T$). Hybrid models are ideally transferred as early as possible, as the final results suffer from geometric inaccuracies developed in the FEM phase due to mesh distortion (Sordo et al., 2023). However, liquefaction failure relies on FEM to capture the effects of earthquake shaking and its associated pore pressure response. Thus, a proper transfer time is one before severe mesh distortion but after the main shaking and development of liquefaction. A transfer time of $t_T = 16$ s is selected for this analysis because by this time the main shaking has occurred, a failure mass has clearly developed (Figure 4), and excess pore pressures have accumulated (Figure 5), yet the mesh is only mildly distorted.

For the MPM phase it is critical to distinguish between unliquefied and liquefied tailings in a manner consistent with the pore pressure response results of our FEM phase (Figure 5). The CB-Geo MPM code, at present, lacks the complex constitutive models available in OpenSees, including PM4Sand. Therefore, the tailings in the MPM phase are modeled with the Mohr-Coulomb constitutive model with $\phi = 0$ and c = $s_R$. As the $s_R$ of the liquefied tailings varies with depth, it is modeled by several different materials. In the transfer script, the tailings material points are assigned a $s_R$ if they have reached an $r_u$ greater than 0.7 or exist at a depth of 12 m or less. This region is indicated in Figure 5.

## 6 MPM PHASE OF THE MOCHIKOSHI MODEL

The MPM phase of the hybrid model simulates the runout of the failure and delivers its final extent. It is initiated by the geometry, stresses, and velocities of the FEM phase at $t = 16.0$ s. The embankment material is modeled via a single Mohr-Coulomb material with c' = 25 kPa and $\phi'= 35°$. The tailings are divided into six different Mohr-Coulomb materials. The unliquefied tailings are assigned c' = 0 and $\phi'= 34°$, matching the values measured by Ishihara (1984) and used by Byrne and Seid-Karbasi (2003). The liquefied tailings are divided into five materials with different c = $s_R$ based on depth, ranging from 3.2 kPa to 9.1 kPa, per the effective stress to $s_R$ relationship of Weber (2015). The foundation soil is modeled as linear-elastic as it is not expected to experience any significant shearing.

To allow the model to run out fully, we include a frictional boundary extending downstream of the dam with a coefficient of friction of 0.35.

Figure 6 shows displacement amplitudes from the FEM model at the transfer time and of the MPM phase at its end. Approximately 5 m of displacement occurs in the FEM phase, after which the model continues displacing downslope for approximately another 10 m in the MPM phase. The upper dam swells, slumps, and settles downstream without fully breaching. Thus, the model does not capture the catastrophic runout of the actual dam failure.

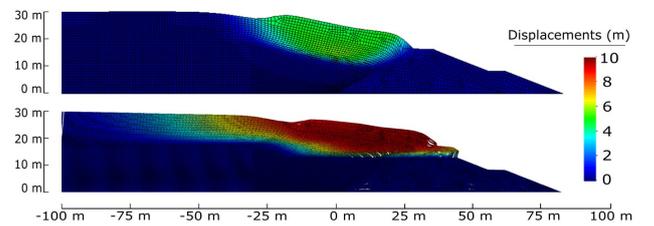

Fig 6: Displacements at the end of the FEM stage (top) and additional displacements from the MPM stage (bottom).

There are multiple components in the hybrid analysis that could be contributing to the limited runout. Our estimates of the residual strength may be too large because they are based on the initial effective stress for the original geometry (before movements), and it is unclear how the residual strength will change as the material flows during the MPM phase. Furthermore, the model is two-dimensional, which underestimates the mass of tailings bearing on the dam and flowing out, as shown in Figure 1. Overall, while we successfully created a hybrid FEM-MPM model of the Mochikoshi Tailings Dam failure, the sequential coupling of these two methods requires that both the FEM and MPM phases be properly calibrated. The MPM phase of the model, particularly the strength and mass components, requires further adjustment to accurately capture the runout behavior. Further research is planned that will explore these issues, the sensitivity of the results to different input parameters, and their ability to captured the observed response in the field.

## CONCLUSIONS

This paper used a hybrid FEM-MPM analysis to model the Mochikoshi Tailings Dam failure. The FEM phase captures the effects of the earthquake on the earth structure: the inertial seismic loading, the development of pore pressure, and the triggering of liquefaction. This dynamic response triggers a failure, and once it has begun, the state of the FEM model is transferred into an MPM model. The MPM phase simulates the post-liquefaction, large deformation behavior of the tailings dam. This initial application of the hybrid FEM-MPM analysis to an earthquake-induced tailings dam failure accurately captured the triggering of the failure and the initial large deformation response, but it did not capture the observed massive runout of hundreds of meters. The MPM phase therefore requires further adjustment and refinement to accurately replicate the large runout behavior of this tailings dam failure.